\documentclass[12pt]{article}
\usepackage{graphicx}
\begin{document}
\rightline{NKU-04-SF2}
\bigskip
\begin{center}
{\Large\bf Gravitational perturbation and quasi-normal modes  of  charged  black holes in Einstein-Born-Infeld gravity}

\end{center}
\hspace{0.4cm}
\begin{center}
Sharmanthie Fernando \footnote{fernando@nku.edu} \\
{\small\it Department of Physics \& Geology}\\
{\small\it Northern Kentucky University}\\
{\small\it Highland Heights}\\
{\small\it Kentucky 41099}\\
{\small\it U.S.A.}\\

\end{center}

\begin{center}
{\bf Abstract}
\end{center}

\hspace{0.7cm}

Born-Infeld electrodynamics has attracted considerable interest due to its relation to strings and D-branes. In this paper  gravitational  
perturbations  of  electrically charged black holes in Einstein-Born-Infeld gravity are studied. The effective potentials for axial perturbations are derived and discussed. The quasi normal modes for the gravitational perturbations are computed using the WKB method. The modes are  compared with those of the Reissner-Nordstrom black hole.  The relation of the quasi normal modes with the non-linear parameter and the spherical index are also investigated. Comments on stability of the black hole and on future directions are made.\\



\section{Introduction}

Born and Infeld developed a non-linear theory of electrodynamics in 1930 to obtain a finite energy model for electron \cite{born}. Born-Infeld theory has attracted lot of interest  lately since it emerges naturally in certain string inspired models \cite{leigh}. The low energy effective action for an open superstring lead to Born-Infeld type action \cite{frad}. Born-Infeld action also arises as an effective action for vector fields on D-branes \cite{tsey}. A good review for Born-Infeld theory can be found in \cite{gib1}.

In this paper  we study   gravitational perturbations and the quasi-normal modes  of static charged black hole in  Einstein-Born-Infeld gravity. The properties of this black hole is discussed in \cite{rasheed} \cite{nora}.

Perturbations of black holes are one of the ways to understand the properties of black holes. Given the fact that black holes are not closed systems, when black holes are perturbed, the resulting modes corresponds to waves with frequencies  with complex numbers: such frequencies depends only on the mass, angular momenta, charge and others parameters of the theory. QNM's are studied extensively in the literature: a good review is given by Kokkotas\cite{kok1}.

 There are several reasons to study QNM's of black holes: one of the reason is its appearance in Loop Quantum Gravity. In Loop Quantum Gravity,  high overtone frequencies are related to the  Barberr-Immirizi parameter $\gamma$ which has sparked many works to  compute high overtone QNM frequencies \cite{hod} \cite{corichi} \cite{mot1} \cite{dreyer} \cite{mot2} \cite{van}  \cite{set1} \cite{set2} \cite{kun} \cite{ander}.

A second reason for studying  QNM to become popular is the AdS/CFT  correspondence \cite{aha}.  QNM frequencies of the black holes in AdS space are related to the  time scale of the CFT on the boundary to reach thermal equilibrium. There are  many work on AdS black holes  on this subject such as \cite{chan}\cite{horo} \cite{car1} \cite{car2}\cite{moss} \cite{wang}  \cite{kok2} \cite{kon1} \cite{kon2} \cite{kon3} \cite{kon4} \cite{kon5} \cite{li} \cite{aros} \cite{car3} \cite{abd}.

The paper is presented as follows: In section 2 the black hole solutions are introduced. In section 3 the gravitational perturbations are given. In section 4 the QNM's are computed and discussed. Finally, the conclusion is given in section 5.


\section{Static charged  black hole in Einstein-Born-Infeld gravity}

Here, we introduce the action for Born-Infeld electrodynamics. First, the action is given by,

\begin{equation}
S_{BI} = \int d^4x \sqrt{-g} \left[ \frac{R }{16 \pi G} + L(F) \right]
\end{equation}
Here $L(F)$ is,
\begin{equation}
L(F) = 4 \beta^2 \left( 1 - \sqrt{ 1 + \frac{ F^{\mu \nu}F_{\mu \nu}}{ 2 \beta^2}} \right)
\end{equation}
In this expression, $\beta$ which is known as the Born-Infeld parameter has dimensions $length^{-2}$ and $G$  has  $length^2$. In the following sections  we will  assume that $16 \pi G = 1$. Note that when $\beta \rightarrow \infty$, the Lagrangian $L(F)$ approaches the Lagrangian  for Maxwell's  electrodynamics given by $ - F^2$. In  the  weak field limit, $L(F)$ has to be of the form 
\begin{equation}
L(F) = - F^{\mu \nu}F_{\mu \nu} + O(F^4)
\end{equation}
\noindent
To describe the equations of motion for the electrodynamic field strength, a second rank tensor $G^{\mu \nu}$ is defined as,
\begin{equation}
G^{\mu \nu} = \frac{F^{\mu \nu}}{\sqrt{1 + \frac{F^2}{2 \beta^2} } }
\end{equation}
By extremising $L(F)$, with respect to $A_{\mu}$, one obtain the equations of motion,
\begin{equation}
\bigtriangledown_{\mu} G^{\mu \nu} = 0 \Rightarrow d * G =0
\end{equation}
The field strength also satisfy the Bianchi identity,
\begin{equation}
\bigtriangledown_{[\mu} F_{ \nu \theta ]} =  0 \Rightarrow d F =0
\end{equation}
Note that the Bianchi identity is satisfied since the field strength can be written as $F_{\mu \nu} = \partial_{\mu} A_{\nu} - \partial_{\nu} A_{\mu}$. On the other hand, $G_{\mu \nu}$ does not satisfy the Bianchi identity.
In comparison, the equations for Maxwell electrodynamics are,
\begin{equation}
\bigtriangledown_{\mu} F^{\mu \nu} = 0 \Rightarrow d * F =0 \hspace{0.5cm} \mbox{and} \hspace{0.5cm} \bigtriangledown_{[\mu} F_{ \nu \theta ]} = 0 \Rightarrow d F =0
\end{equation}
For static spherically symmetric case,
\begin{equation}
G_{tr} = - \frac{Q}{r^2}  \hspace{0.5cm} \mbox{or}\hspace{0.5cm}   F_{tr} = E(r) = - \frac{Q}{\sqrt{ r^4 + \frac{Q^2}{\beta^2}}}
\end{equation}
Note that for this case the  non-linear Lagrangian reduces to,
\begin{equation}
 L(F) = 4 \beta^2 \left( 1 - \sqrt{ 1 - \frac{E^2}{\beta^2}} \right)
\end{equation} 
imposing  an upper bound for $|E|$ $\leq$ $\beta$. This is a vital characteristic of Born-Infeld electrodynamics which leads to finite self energy of the electron.

By varying the total Lagrangian in eq.(1) with respect to $g_{\mu \nu}$ and assuming $16 \pi G=1$, the gravitational field equations are obtained as,
\begin{equation}
R_{\mu \nu} =  T_{\mu \nu}
\end{equation}
where,
\begin{equation}
T_{\mu \nu} = - 2 \left(  \frac{ g^{\alpha \theta} F_{\mu \alpha}F_{\nu \theta }}{\sqrt{1 + \frac{F^2}{2\beta^2}}} +   \frac{ g_{\mu \nu} L(F)}{4}   \right)
\end{equation}
Note that for $\beta \rightarrow \infty$, $T_{\mu \nu}$ approaches the energy-momentum tensor for the Maxwell's electrodynamics given by,
\begin{equation}
T_{\mu \nu} = -2 \left( g^{ \alpha \theta} F_{\mu \alpha} F_{\nu \theta} - \frac{ g_{\mu \nu}}{4}  F^2 \right)
\end{equation}
By solving the field equations, the static charged black hole with spherical symmetry can be obtained as,
\begin{equation}
ds^2 = f(r) dt^2 - f(r)^{-1} dr^2 - r^2 ( d \theta^2 + Sin^2(\theta) d \varphi^2)
\end{equation}
with,
\begin{equation}
f(r) = 1 - \frac{2M}{r} + 2 \beta \left( \frac{r^2 \beta}{3} - \frac{1}{r} \int_{r}^{\infty}  \sqrt{ Q^2 + r^4 \beta^2} \right)
\end{equation}
In the limit $\beta \rightarrow \infty$, the elliptic integral  is  expanded to give,
\begin{equation}
f(r) = 1 - \frac{2 M}{r} + \frac{ Q^2}{r^2}
\end{equation}
resulting  the function $f(r)$ for the Reissner-Nordstrom black hole which arises from gravity coupled to  Maxwell's electrodynamics.

The Hawking temperature of the black hole is given by,
\begin{equation}
T_H = \frac{1}{4\pi} \left[ \frac{1}{r_{eh}}  + 2\beta \left( r_{eh} \beta - \frac{\sqrt{(Q^2 + r_{eh}^2 \beta^2)}}{r_{eh}} \right) \right]
\end{equation}
Here, $r_{eh}$ is the event horizon of the black hole which is a solution of $f(r)=0$. The above black hole and its thermodynamic properties are discussed in \cite{rasheed} and static charged black hole solution to Born-Infeld gravity with a cosmological constant was presented in \cite{fer1} and was extended to higher dimensions in \cite{dey}.


\section{Metric perturbations}

In this section the equations for gravitational perturbations of charged black holes in Born-Infeld gravity are given. The notations of \cite{chandra} are closely followed. 

As described in section(2), Born-Infeld black hole is a stationary spherically symmetric time-independent solution of the field equations. Once perturbed, the metric will become non-stationary and time-dependent. Here, the perturbed metric will be assumed to be axially symmetric. The most general metric which is time-dependent as well as axially symmetric can be written as,
\begin{equation}
ds^2 = e^{ 2 \nu} dt^2 - e^{2 \psi} ( d \varphi - q_2 dx^2 - q_3 dx^3 - \omega dt)^2 - e^{ 2 \mu_2} (dx^2)^2 - e^{ 2 \mu_3} (dx^3)^2
\end{equation}
Here, $\nu, \psi, \mu_2, \mu_3, \omega, q_2, q_3$ are functions of $t,x^2,x^3$ only. These functions do not depend on $\varphi$ preserving axisymmetry.

The charged Born-Infeld black hole discussed in the previous section will be considered as a special, spherically symmetric time-independent solution of the above line element.
Hence the unpertubed metric consists of the following functions;
\begin{equation}
e^{ 2 \nu} = e^{ - 2 \mu_2} =  1 - \frac{2M}{r} + 2 \beta \left( \frac{r^2 \beta}{3} - \frac{1}{r} \int_{r}^{\infty}  \sqrt{ Q^2 + r^4 \beta^2} \right)
 = \frac{\bigtriangleup}{r^2}
\end{equation}
and
\begin{equation}
e^{\mu_3} = r, \hspace{0.5 cm} e^{\psi} = r Sin \theta, \hspace{0.5 cm} \omega= q_2 = q_3  = 0, \hspace{0.5cm} \bigtriangleup = r^2 e^{2 \nu}
\end{equation}
with the coordinates $x^2 = r$ and $x^3= \theta$.
Also the electrodynamic field strength $F_{ab}$  and the Ricci tensor for the unperturbed metric has the following values;
\begin{equation}
F_{02} = -\frac{Q}{r^2}
\end{equation}
\begin{equation}
R_{00} = -R_{22} = -2 \beta^2 \left( 1 - \frac{1}{ \sqrt{ 1 -\frac{F_{02}^2}{\beta^2} } } \right)
\end{equation}
\begin{equation}
R_{11} = R_{33} = 2 \beta^2 \left( 1 - \sqrt{ 1 - \frac{ F_{02}^2}{ \beta^2} } \right)
\end{equation}
\begin{equation}
R_{01} = R_{02} = R_{03} = R_{12} = R_{13} = R_{23} = 0
\end{equation}

A general perturbation of the black hole will lead to non zero values of $\omega, q_2, q_3$, $\delta \nu, \delta \mu_2, \delta \mu_3$ and $\delta \psi$. The perturbations leading to non-vanishing values of $\omega, q_2$ and $q_3$ are called axial perturbations and the perturbations leading to increments to $\nu, \mu_2, \mu_3$, $\psi$ are called polar perturbations. This will be discussed in detail in the next section. The perturbed equations are obtained by linearising the Einstein and Born-Infeld's equation around the unperturbed metric given in eq.(13).

To obtain the linearised perturbation equations for the above black hole, it is convenient to use the tetrad formalism to the space-time metric in eq.(17). We will use the indices $a,b=0,1,2,3$ for the  orthonormal basis and $\mu, \nu = 0,1,2,3$ for the coordinate basis. To handle the black hole in this paper the coordinates are chosen as $x^0 = t, x^1 = \varphi, x^2 = r$ and $ x^3=\theta$. The tetrad basis $e^{a}_{\mu}$ is chosen such that $e^a_{\mu} e^{b}_{\nu} \eta_{ab} = g_{\mu \nu}$, where $\eta_{ab} = (1,-1,-1,-1)$. For the metric in eq.(17), the tetrads $e^{a}_{\mu}$ are given by,
\begin{equation}
e_{\mu}^{(0)} = ( e^{\nu}, 0,0,0 )
\end{equation}
\begin{equation}
e_{\mu}^{(1)} = ( -\omega e^{\psi},  e^{\psi}, -q_2 e^{\psi}, - q_3 e^{\psi} )
\end{equation}
\begin{equation}
e_{\mu}^{(2)} = ( 0,0, e^{ \mu_2}, 0 )
\end{equation}
 \begin{equation}
 e_{\mu}^{(3)} = ( 0,0,0, e^{\mu_3} )
\end{equation}
Tensors in  coordinate and orthonormal basis are related to each other by the tetrads. For example the field strength $F_{\mu \nu} = F_{ab} e^a_{\mu} e^{b}_{\nu}$.

\subsection{Born-Infeld equations}

As mentioned in the previous section, there are total of eight equations resulting from the Bianchi identities and the equations of motion for Born-Infeld electrodynamics. The four equations resulting from Bianchi identities  $\bigtriangledown_{[\mu} F_{ \nu \theta ]} = 0$ are as follows,
\begin{equation}
\left( e^{ \psi + \mu_2} F_{12} \right)_{,3} + \left( e^{ \psi + \mu_2} F_{31} \right)_{, 2} =0
\end{equation}
\begin{equation}
\left( e^{ \psi + \nu} F_{01} \right)_{,2} + \left( e^{ \psi + \mu_2} F_{12} \right)_{, 0} =0
\end{equation}
\begin{equation}
\left( e^{ \psi + \nu} F_{01} \right)_{,3} + \left( e^{ \psi + \mu_3} F_{13} \right)_{, 0} =0
\end{equation}
$$
\left( e^{ \nu + \mu_2} F_{02} \right)_{,3} - \left( e^{ \nu + \mu_3} F_{03} \right)_{, 2} + \left( e^{\mu_3 + \mu_2} F_{23}  \right)_{,0}$$ 
\begin{equation}
=  e^{\psi + \nu} F_{01} Q_{23} + e^{\psi + \mu_2} F_{12} Q_{03} - e^{ \psi + \mu_3} F_{13} Q_{02}
\end{equation}
There are four equations resulting from $\bigtriangledown_{\mu} G^{\mu \nu}=0$ in the orthonormal basis as follows,
\begin{equation}
\left( e^{ \psi + \mu_3} G_{02} \right)_{,2} + \left( e^{ \psi + \mu_2} G_{03} \right)_{, 3} =0
\end{equation}
\begin{equation}
- \left( e^{ \psi + \mu_3} G_{23} \right)_{,2} + \left( e^{ \psi + \mu_2} G_{03} \right)_{, 0} =0
\end{equation}
\begin{equation}
\left( e^{ \psi + \nu} G_{23} \right)_{,3} + \left( e^{ \psi + \mu_3} G_{02} \right)_{, 0} =0
\end{equation}
$$
\left( e^{ \mu_2 + \mu_3} G_{01} \right)_{,0} + \left( e^{ \nu + \mu_3} G_{12} \right)_{, 2} + \left( e^{ \nu + \mu_2} G_{13} \right)_{, 3}$$
\begin{equation}
= e^{ \psi + \mu_2} G_{02} Q_{02} + e^{ \psi + \mu_2} G_{03} Q_{03} - e^{ \psi + \nu} G_{23} Q_{23}
\end{equation}
Here, the partial derivative of a function $g$ is given with the notation, 
\begin{equation}
( g )_{,a} =  \frac{\partial g}{\partial x^a}
\end{equation}
The function $Q_{AB}$ are given by,
\begin{equation}
Q_{A0}  = \frac{\partial q_A}{\partial x^0} - \frac{\partial \omega}{\partial x^A}
\hspace{0.5cm}
\mbox{and}\hspace{0.5cm}
Q_{AB} = \frac{\partial q_A}{\partial x^B} - \frac{\partial q_B}{\partial x^A}
\hspace{0.5cm} ( A,B = 2,3)
\end{equation}
In the two groups of equations,  eq.(28) and eq.(32) can be ignored since they just provide integrability conditions for the two following equations.

In the orthonormal basis, the only non-zero components of $G_{ab}$ and $F_{ab}$ for the unperturbed metric are,
\begin{equation}
G_{02} = -\frac{Q}{r^2}
\end{equation}
and
\begin{equation}
F_{02} = -\frac {Q}{\sqrt{r^4 + Q^2/\beta^2}}
\end{equation}
Equations (29), (30) and (35) can be written in  the spherical coordinates adopted for the black hole as,
\begin{equation}
\left( r e^{\nu} F_{01} Sin \theta \right)_{,r} +  r e^{- \nu} F_{12,t} Sin \theta = 0
\end{equation}
\begin{equation}
r e^{\nu} \left( F_{01} Sin \theta \right)_{, \theta} + r^2 F_{13,t} Sin \theta = 0
\end{equation}
\begin{equation}
r e^{- \nu} G_{01,t} + \left( r e^{\nu} G_{12} \right)_{,r} + G_{13,\theta} = - Q ( \omega_{,2} - q_{2,0})Sin \theta 
\end{equation}
Also the equations (33), (34) and (31) written in spherical coordinates simplifies to,
\begin{equation}
r e^{-\nu} G_{03,t} = \left( r e^{\nu} G_{23} \right)_{,r}
\end{equation}
\begin{equation}
\delta G_{02,t} - \frac{Q}{r^2} \left( \delta \psi + \delta \mu_3 \right)_{,t} + \frac{e^{\nu}}{r Sin \theta } \left( G_{23} Sin \theta \right)_{, \theta} = 0
\end{equation}
\begin{equation}
\left[ \delta F_{02} - \frac{Q}{r^2} \left( \delta \nu + \delta \mu_2 \right) \right]_{,\theta} +  ( r e^{\nu} F_{ 30} )_{,r} + r e^{- \nu} F_{23, t} = 0
\end{equation}

\subsection{The perturbation of the Ricci tensor}

To facilitate the explanation of the equations, the Ricci tensor components for the general metric in eq.(17) in the orthonormal basis is given here. Note that these are given in Chandrasekhar's book \cite{chandra}.

$$R_{00} = - e^{ - 2 \nu} \left[ ( \psi + \mu_2 + \mu_3)_{,0,0} + \psi_{,0} ( \psi - \nu)_{,0} + \mu_{2,0} ( \mu_2 - \nu)_{,0} + \mu_{3,0} ( \mu_3 - \nu)_{,0} \right]$$
$$+ e^{ - 2 \mu_2} \left[ \nu_{,2,2} + \nu_{,2} ( \psi + \nu - \mu_2 + \mu_3)_{,2} \right] + e^{- 2 \mu_3} \left[ \nu_{,3,3} + \nu_{,3} ( \psi + \nu + \mu_2 - \mu_3)_{,3} \right]$$
\begin{equation}
- \frac{1}{2} e^{ 2 \psi - 2 \nu} \left[ e^{-2 \mu_2} Q_{20}^2 + e^{-2 \mu_3} Q_{30}^2 \right]
\end{equation}
$$ R_{11} = - e^{-2 \mu_2} \left[ \psi_{,2,2} + \psi_{,2} ( \psi + \nu + \mu_3 - \mu_2)_{,2} \right] - e^{- 2 \mu_3} \left[ \psi_{,3,3} + \psi_{,3} ( \psi + \nu + \mu_2 - \mu_3)_{,3} \right]$$
$$ + e^{ - 2 \nu} \left[ \psi_{,0,0} + \psi_{,0} ( \psi - \nu + \mu_2 + \mu_3)_{,0} \right] + \frac{1}{2} e^{ 2 \psi - 2 \mu_2 - 2 \mu_3} Q_{23}^2$$
\begin{equation}
- \frac{1}{2} e^{ 2 \psi - 2 \nu} \left[ e^{-2 \mu_3} Q_{30}^2 + e^{-2 \mu_2} Q_{20}^2 \right]
\end{equation}
$$ R_{22} = - e^{ - 2 \mu_2} \left[ ( \psi + \nu+ \mu_3)_{,2,2} + \psi_{,2} ( \psi - \mu_2)_{,2} - \mu_{3,2} ( \mu_3 - \mu_2)_{,2} + \nu_{,2} ( \nu - \mu_2)_{,2} \right]$$
$$ - e^{- 2 \mu_3} \left[\mu_{2,3,3}   + \mu_{2,3}( \psi + \nu +\mu_2 - \mu_3)_{,3} \right] + e^{- 2 \nu} \left[ \mu_{2,0,0} + \mu_{2,0} ( \psi - \nu + \mu_2 + \mu_3)_{,0} \right]$$
\begin{equation}
- \frac{1}{2} e^{ 2 \psi - 2 \mu_2} \left[ e^{- 2 \mu_3} Q_{23}^2 - e^{-2 \nu} Q_{20}^2 \right]
\end{equation}
\begin{equation}
R_{01} = - \frac{1}{2} e^{ - 2 \psi - \mu_2 - \mu_3} \left[ ( e^{3 \psi - \nu - \mu_2 + \mu_3 } Q_{20} )_{,2} + ( e^{3 \psi - \nu - \mu_3 + \mu_2 } Q_{30} )_{,3} \right]
\end{equation}
\begin{equation}
R_{12} = -\frac{1}{2} e^{- 2 \psi - \nu + \mu_3} \left[ \left( e^{3 \psi + \nu - \mu_2 + \mu_3} Q_{32} \right)_{,3} + \left( e^{ 3 \psi - \nu + \mu_3 - \mu_2} Q_{02} \right)_{,0} \right]
\end{equation}
$$R_{02} = - e^{ - \mu_2 -\nu} \left[ ( \psi + \mu_3 )_{,2,0} + \psi_{,2} ( \psi - \mu_2)_{,0} + \mu_{3,2} ( \mu_{3} - \mu_2)_{,0} - (\psi + \mu_3)_{,0} \nu_{,2} ) \right] $$
\begin{equation}+ \frac{1}{2} e^{ 2 \psi - \nu - 2 \mu_3 - \mu_2 } Q_{23} Q_{30}
\end{equation}
$$R_{23} = - e^{ - \mu_2 - \mu_3} \left[ ( \psi + \nu)_{,2,3} -\mu_{2,3} (\psi +  \nu)_{,2}  - \mu_{3,2}(\psi + \nu)_{,3}   + \psi_{,2} \psi_{,3} + \nu_{,2} \nu_{,3} \right]$$
\begin{equation}
+ \frac{1}{2} e^{ 2 \psi - 2 \nu - \mu_2 - \mu_3} Q_{20} Q_{30}
\end{equation}
The other components $R_{33}$, $R_{13}$ and $R_{03}$ are not given here. They can be obtained by interchanging the indices 2 and 3 in $R_{22}$, $R_{12}$ and $R_{02}$.
The Ricci tensor for the Born-Infeld electrodynamics is given by,
\begin{equation}
R_{ab} =  -2 \left[ \eta^{cd} \frac{F_{ac} F_{bd} }{\sqrt{ 1 + \frac{F^2}{ 2 \beta^2}}} + \eta_{ab} \left\{ \beta^2 \left( 1 - \sqrt{ 1 + \frac{ F^2}{ 2 \beta^2}} \right) \right\} \right]
\end{equation}
Since  the expressions for the perturbed Ricci tensor is given in terms of the metric functions, one has to compute the changes to the Ricci tensor via the energy momentum tensor to obtain the complete equations. Therefore the perturbed components of the Ricci tensor for the Born-Infeld case are computed as follows;
\begin{equation}
\delta R_{ab} = - 2 \left[ \frac{ 4 F_{02} \delta F_{02} } {\sqrt{1 - \frac{F_{02}^2}{ \beta^2}}} \left( \frac{\eta_{ab}}{4} + \frac{  \eta^{nm} F_{an} F_{bm} }{ 4 \beta^2 ( 1 + \frac{F^2}{ 2 \beta^2} ) } \right) + \eta^{nm} \left( \frac{\delta F_{an} F_{bm} + F_{an} \delta F_{bm} }
{ \sqrt{ 1 + \frac{F^2}{ 2 \beta^2} } }  \right) \right]
\end{equation}
Considering the fact that  only non-zero component of $F_{ab}$  before the perturbation is $F_{02}$, the exact expressions for $\delta R_{ab}$ can be computed as,
\begin{equation}
\delta R_{00} = - \delta R_{22} = - \frac{2 Q}{r^2} 
\frac{ \delta F_{02}}{ ( 1 + \frac{F^2}{2 \beta^2}) }
\end{equation}
\begin{equation}
\delta R_{11} = \delta R_{33} = - \frac{2 Q}{r^2} \delta F_{02}
\end{equation}
\begin{equation}
\delta R_{01} = - \frac{2 Q}{r^2} \delta F_{12}, \hspace{0.5 cm} \delta R_{03} = \frac{ 2 Q}{r^2} \delta F_{23}
\end{equation}
\begin{equation}
\delta R_{12} = \frac{2 Q}{r^2} \delta F_{01}, \hspace{0.5cm} \delta R_{23}= \frac{2 Q}{r^2} \delta F_{03}
\end{equation}
\begin{equation}
\delta R_{13} = \delta R_{02} = 0
\end{equation}


\subsection{Two categories of metric  perturbations}

The metric perturbations will lead to non-zero values of $\delta F_{02}, F_{03}, F_{23}, \delta \nu, \delta \psi, \delta \mu_2 , \delta \mu_3$ and $\omega, q_2, q_3, F_{01}, F_{12}, F_{13}$. If only the terms in first order in perturbations are kept in the equations of motion,  the above two groups of increments  can be treated separately leading to two kinds of perturbations. To clarify this further, one can consider the perturbed Ricci tensors and make the analogy as follows; The components $R_{00}, R_{11}, R_{22}, R_{33}$ will  change only if $\delta F_{02}$ is non-zero. The components $R_{03}$ and $R_{23}$ will undergo changes only if $F_{23}$ and $F_{03}$ are non-zero respectively. By studying the expressions for the Ricci tensors in terms of the metric components, it is clear that this also means  $\delta \nu, \delta \psi, \delta \mu_2, \delta \mu_3$ values has to be non-zero. On the other hand, the components $R_{01}$, $R_{12}$ and $R_{13}$ will  change only if $F_{12}, F_{01}, F_{13}$ are non-zero respectively also meaning $\omega, q_2, q_3$ has to be non-zero. As explained in \cite{chandra}, the first group of rotations impart a rotation to the black hole while the second group does not. Hence the first is called $\it{polar}$ and the second is called $\it{axial}$ perturbations.


\subsection{Axial perturbation}

In this paper only the axial perturbations are considered. It is characterized by
non-zero values of $\omega, q_2, q_3, F_{01}, F_{12}, F_{13}$. The two equations governing axial perturbations comes from $R_{12}$ and $R_{13}$. Note that before perturbations, $R_{12}= R_{13}=0$. After perturbations $\delta R_{13}=0$ and $\delta R_{12} \neq 0 $. By substituting the changes in the corresponding Ricci tensors and keeping the functions  $\nu, \psi, \mu_2, \mu_3$ as same as before the perturbations, we obtain the following equations,
\begin{equation}
\left( r^2 e^{ 2 \nu} Q_{23} Sin^3 \theta \right)_{, \theta} + r^4 Q_{02,t} Sin^3 \theta = 2 ( r^3 e^{\nu} Sin^2 \theta ) \delta R_{12} = 4 Q r e^{\nu} F_{01} Sin^2 \theta
\end{equation}
and,
\begin{equation}
\left(r^2 e^{2 \nu} Q_{23} Sin^3 \theta \right)_{,r} - r^2 e^{-2 \nu} Q_{03,t} Sin^3 \theta = - 2 (r^2 Sin^2 \theta) \delta R_{13}=0
\end{equation}
Two new functions are defined as, 
\begin{equation}
F_{01} Sin \theta = B, \hspace{0.5cm} G_{01} Sin \theta = \hat{B}
\end{equation}
Considering the relation in eq.(4), $B$ and $\hat{B}$ are related by $ \hat{B} = B/p$ where,
\begin{equation}
p = \sqrt{ 1 + \frac{ F^2}{2 \beta^2}}
\end{equation}
Eliminating $G_{12}$ and $G_{13}$ from eq.(42) with the help of  eq.(40) and eq.(41) leads to,
\begin{equation}
\left[ \frac{e^{ 2 \nu}}{p} \left( r e^{\nu} \hat{B} p \right)_{,r} \right]_{,r} + 
\frac{ e^{\nu}}{r} \left( \frac{\hat{B}_{, \theta}}{Sin \theta} \right)_{, \theta} Sin \theta - r e^{- \nu} \hat{B}_{,t,t} = Q ( \omega_{,2,0} - q_{2,0,0} ) Sin^2 \theta
\end{equation}
By defining new functions $\hat{Q}(r,\theta,t)$ as,
\begin{equation}
\hat{Q}( r , \theta,t) = r^2 e^{ 2 \nu} Q_{23} Sin^2 \theta = \bigtriangleup ( q_{2,3} - q_{3,2} ) Sin^3 \theta
\end{equation}
eq.(60) and eq.(61) can be re written as,
\begin{equation}
\frac{1}{ r^4 Sin^3 \theta}  \frac{ \partial \hat{Q}}{ \partial \theta} = - ( \omega_{,2} - q_{2,0} )_{,t} + \frac{ 4 Q \hat{B} p e^{ \nu}}{r^3 Sin^2 \theta}
\end{equation}
and
\begin{equation}
\frac{\bigtriangleup}{ r^4 Sin^3 \theta} \frac{ \partial \hat{Q}}{ \partial r} = ( \omega_{,3} - q_{3,0} )_{,t} 
\end{equation}
Taking the time dependence of the perturbed values of $\omega, q_2, q_3, \hat{Q}$ to be  $e^{-i \omega t}$ where $\omega$ is the frequency of the modes,  eq.(66) and eq.(67) are combined to be,
\begin{equation}
r^4 \frac{\partial}{\partial r} \left( \frac{\bigtriangleup}{r^4} \frac{\partial \hat{Q}}{ \partial r} \right) + Sin^3 \theta \frac{\partial}{\partial \theta} \left( \frac{1}{Sin^3 \theta} \frac{\partial \hat{Q}}{\partial \theta} \right) + \frac{ \omega^2 r^4 \hat{Q} }{\bigtriangleup} = 4 Q e^{\nu} r \left( \frac{ \hat{B} p}{Sin^2 \theta} \right)_{,\theta} Sin^3 \theta 
\end{equation}
Combining eq.(64) and eq.(66) leads to,
\begin{equation}
\left[ \frac{e^{2 \nu}}{p} ( r e^{\nu} \hat{B} p )_{,r} \right]_{,r} +
\frac{e^{\nu}}{r} \left( \frac{ \hat{B}_{, \theta}}{Sin \theta} \right)_{, \theta} Sin \theta +
\left( \omega^2 r e^{- \nu} - \frac{ 4 Q e^{\nu} p}{r^3} \right) \hat{B} = - Q \frac{\hat{Q}_{, \theta}} {r^4 Sin \theta}
\end{equation}
An {\it ansatz} similar to the one given for the Reissner-Nordstrom black hole perturbation in \cite{chandra}  for the functions $\hat{Q}$ and $\hat{B}$ are made as,
\begin{equation}
\hat{Q}(r,\theta) = \hat{Q}(r) C_{l+2}^{-\frac{3}{2}} (\theta)
\end{equation}
\begin{equation}
\hat{B}(r,\theta) = \frac{\hat{B}(r)}{Sin \theta} \frac{ d C_{l+2}^{-\frac{3}{2}} } { d \theta} = 3 \hat{B}(r) C_{l+1}^{-\frac{1}{2}}(\theta)
\end{equation}
Here, $C_{n}^{\nu}$ denotes the Gegenbauer function which is a solution to the following differential equation,
\begin{equation}
\left[ \frac{d}{d \theta} Sin^{ 2 \nu} \theta \frac{d}{d \theta} + n(n + 2 \nu) Sin^{2 \nu} \theta \right] C_{n}^{\nu} ( \theta) = 0
\end{equation}
Gegenbauer function also satisfy the following recurrence relation,
\begin{equation}
\frac{1}{Sin \theta} \frac{d C_{\nu}^{n}} { d \theta} = - 2 \nu C_{n-1}^{\nu +1}
\end{equation}
By substituting the functions $\hat{Q}(r,\theta)$ and $\hat{B}(r,\theta)$ in eq.(68) and eq.(69), the following two equations are obtained,
\begin{equation}
\bigtriangleup \frac{d}{dr} \left( \frac{ \bigtriangleup}{r^4} \frac{ d \hat{Q}}{dr} \right) - \mu^2 \frac{ \bigtriangleup}{r^4} \hat{Q} + \omega^2 \hat{Q} 
= - \frac{ 4 Q}{ r^3} \mu^2 \bigtriangleup e^{\nu} \hat{B} p
\end{equation}
and
\begin{equation}
\left[ \frac{ e^{2 \nu}}{ p} \left( r e^{\nu} \hat{B} p \right)_{,r} \right]_{,r} - ( \mu^2 + 2) \frac{ e^{\nu} }{r} \hat{B}  + \left( \omega^2 r e^{-\nu} - \frac{4 Q}{r^3} e^{\nu} p \right) \hat{B} = - Q  \frac{\hat{Q}}{r^4}
\end{equation}
Here,
\begin{equation}
\mu^2 = 2 n = (l-1) ( l+2 )
\end{equation}
The above eq.(74) and (75) are the main equations for the axial perturbations.
The variable $r$ will be replaced in terms of the ``tortoise'' coordinate defined by,
\begin{equation}
e^{ 2 \nu} \frac{d}{d r} = \frac{d}{ dr_*}
\end{equation}
$\hat{Q}(r)$ and $\hat{B}(r)$ are redefined in terms of functions $H_1$ and $H_2$ as,
\begin{equation}
\hat{Q} = r H_2, \hspace{0.5cm} r e^{\nu} \hat{B} p = -  \frac{H_1 e^{-\frac{\phi}{2} }}{ 2 \mu}
\end{equation}
Here,
\begin{equation}
e^{\phi} = \frac{1}{p} \Rightarrow  \phi = ln\left[ \frac{\sqrt{Q^2 + r^4 \beta^2}}{r^2 \beta} \right]
\end{equation}
With these definitions, eq.(74) becomes,
\begin{equation}
\Lambda^2 H_2 = \frac{\bigtriangleup}{r^5} \left\{ \left[ \mu^2 r - ( e^{2 \nu} r )' r + 3 e^{2 \nu} r \right] H_2 + 2 Q \mu e^{-\frac{\phi}{2}} H_2 \right\}
\end{equation}
and eq.(75) becomes,
\begin{equation}
\Lambda^2 H_1 = \frac{\bigtriangleup}{r^5} \left\{ \left[ ( \mu^2 +2) r +  
\frac{4 Q^2 e^{-\phi}}{r} + r^3 \left( \frac{ (\phi' e^{2 \nu} )'}{2} + \frac{ (\phi')^2 e^{ 2 \nu}}{4} \right) \right] H_2 + 2 \mu Q e^{- \frac{\phi}{2}} H_1 \right\}
\end{equation}
Here,
\begin{equation}
\Lambda^2 = \frac{d^2}{dr_*^2} + \omega^2
\end{equation}
As $\beta \rightarrow \infty$, both equations approaches the Reissner-Nordstrom black hole expressions given in \cite{chandra}.


\subsection{Effective potentials}

The above two equations can be considered as two one-dimensional wave equations coupled by the interaction matrix,
\begin{equation}
\left(\frac{d^2}{dr_*^2} + \omega^2 \right) \left( \begin{array}{ll}
			H_{1} \\
			H_{2}
			\end{array}
				\right)
 =  \left( \begin{array}{ll}
			U_{11} & U_{12} \\
			U_{21} & U_{22}
			\end{array}
				\right) \left( \begin{array}{ll}
			H_{1} \\
			H_{2}
			\end{array}
				\right)
\end{equation}
where,
\begin{equation}
U_{11}=\frac{\bigtriangleup}{r^5}  \left[ ( \mu^2 +2) r +  
\frac{4 Q^2 e^{-\phi}}{r} + r^3 \left( \frac{ (\phi' e^{2 \nu} )'}{2} + \frac{ (\phi')^2 e^{ 2 \nu}}{4} \right) \right]
\end{equation}
\begin{equation}
U_{12}=U_{21}=2 Q \mu \frac{\bigtriangleup}{r^5} e^{-\frac{\phi}{2}}
\end{equation}
\begin{equation}
U_{22} = \frac{ \bigtriangleup}{r^5} \left[ \mu^2 r - ( e^{2 \nu} r )' r + 3 e^{2 \nu} r \right] 
\end{equation}
The matrix $U$ can be diagonalized by a similarity transformation leading to one-dimensional Schrodinger type wave equation given by,
\begin{equation}
\Lambda^2 Z_i = V_i Z_i \hspace{0.5cm} ( i=1,2) 
\end{equation}
where,
$$V_1 = \frac{1}{2} \left(  U_1 + U_2 + \sqrt{ ( U_1 - U_2)^2 + 4 U_{12}^2 } \right)$$
\begin{equation}
V_2 = \frac{1}{2} \left(  U_1 + U_2 - \sqrt{ ( U_1 - U_2)^2 + 4 U_{12}^2 } \right)
\end{equation}
The perturbation equations of the Schwarzchild black hole were derived by Regge and Wheeler \cite{regge} and Zerilli \cite{zer1}. The equations for the Reissner-Nordstrom black hole were first derived by Moncrief and Zerilli \cite{mon}\cite{zer2}. The Kerr black hole equations were derived by Teukolsky \cite{teuk}. The equations for the charged black hole with a dilaton were derived in \cite{wil} \cite{fera}.

The exact expressions for $V_{1,2}$ are not given here. However, the effective potential $V_1$ for the Born-Infeld black hole is plotted to show how it changes with charge $Q$ and the non-linear parameter $\beta$ in the following figures.

\begin{center}
\scalebox{.9}{\includegraphics{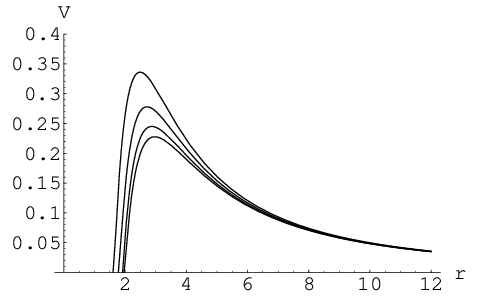}}

\vspace{0.3cm}
\end{center}
Figure 1. The behavior of the effective potential $V_1(r)$ with 
the charge for the Born-Infeld black hole. Here, $M=1$, $\beta=0.2$ and $l=2$. The height of the potential decreases when the charge increases.

\begin{center}
\scalebox{.9}{\includegraphics{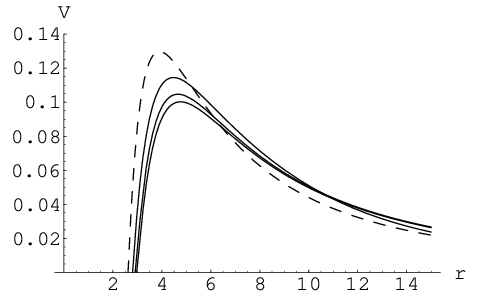}}

\vspace{0.3cm}
\end{center}
Figure 2. The behavior of the effective potential $V_1r)$ with 
the non-linear parameter $\beta$. Here, $M=1.5$, $Q=1$ and $l=2$. The maximum height of the potential increases as $\beta$ increases. The dashed one is the potential for the Reissner-Nordstrom black hole with same mass and charge.


\subsection{Remarks on stability}

The potentials are real and positive out side the event horizon as it is evident from the above figures. Hence following the arguments by Chandrasekhar \cite{chandra} the Born-Infeld black holes can be considered stable classically.


\section{Quasi-normal modes of the Born-Infeld black hole}

As discussed in the introduction, Quasi-normal modes (QNM) arises as a result of perturbations of black hole space-times.  Equations of perturbations lead to wave solutions. Boundary conditions has to be imposed on such equations to obtain QNM frequencies: they are purely ingoing at the horizon and purely out going for asymptotically black hole space-times.

In general the fundamental equation of black hole perturbations given in eq.(87) cannot be solved analytically. We can give few examples which can be solved exactly: in 2+1 dimensions there are two black hole solutions (BTZ black hole \cite{bir1}  and the charged dilaton black hole \cite{fer2}); in five dimensions, exact values are obtained for vector perturbations by Nunez and Starinets \cite{nun}.

There are several techniques   to compute QNM's in literature.  In this paper, a semi analytical technique developed by Iyer and Will \cite{will} is followed. The method makes use of the WKB approximation, carried out to the third  order. This approach has been applied to the Schwarzschild \cite{iyer} Reissner-Nordstrom \cite{koko1}, charged dilaton black hole \cite{fera} \cite{kon5} \cite{fer3}. The basics of this method is reviewed as follows;

Take the perturbation eq.(87) in the following form.
\begin{equation}
\left(\frac{d^2}{dr_{*}^2} + Q (r_{*}) \right)  Z(r_{*}) = 0
\end{equation}
Here $Q(r_*) = \omega^2 - V_{1,2}(r_*) $. Then, one can define new variables $ \Lambda(n), \Omega(n), \hat{\Lambda}(n), \hat{\Omega}(n), \alpha $ as follows.
\begin{equation}
\Lambda(n) = \frac{1}{(2 Q_0^{(2)})^{1/2}} \left[\frac{1}{8} \left[ \frac{Q_0^{(4)}}{Q_0^{(2)}} \right] ( \frac{1}{4} + \alpha^2 ) -\frac{1}{288} \left[\frac{Q_0^{(3)}}{Q_0^{(2)} } \right]^2 ( 7 + 60 \alpha^2 ) \right]
\end{equation}
$$
\Omega(n) = \frac{ n + \frac{1}{2}}{2 Q_0^{(2)}} \left[ \frac{5}{6912} \left[ \frac{ Q_0^{(3)}}{Q_0^{(2)}} \right]^4 ( 77 + 188 \alpha^2) - \frac{1}{384} \left[ \frac{(Q_0^{(3)})^2 Q_0^{(4)} }{ (Q_0^{(2)})^3} \right] ( 51 + 100 \alpha^2) \right.
$$
\begin{equation} 
\left. + \frac{1}{2304} \left[ \frac{Q_0^{(4)}}{Q_0^{(2)}} \right]^2 ( 67 + 68 \alpha^2) + \frac{1}{288} \left[ \frac{(Q_0^{(2)})^3 Q_0^{(5)}}{(Q_0^{(2)})^2} \right] ( 19+ 28 \alpha^2) 
-\frac{1}{288} \left[ \frac{Q_0^{(6)}}{ Q_0^{(2)} } \right] 
( 5 + 4 \alpha^2)\right] 
\end{equation}

\begin{equation}
\alpha= n + \frac{1}{2}; \hspace{1 cm} \hat{\Lambda}(n) = - i \Lambda(n)     ;\hspace{2 cm} \hat{\Omega}(n) = \Omega/(n + \frac{1}{2})
\end{equation}
Note that the superscript $(n)$ denotes the appropriate number of derivatives of $Q(r_*)$ with respect to $r_*$ evaluated at the maximum of $Q(r_*)$. In the case of black hole perturbations where $V(r_*)$ is independent of frequency $\omega$, the quasi normal modes frequencies are given by,
\begin{equation}
\omega^2(n) = [  V_0 + ( -2 V_0^{(2)})^{1/2} \hat{\Lambda}(n)] - i  (n + \frac{1}{2}) ( - 2 V_0^{(2)})^{1/2} [ 1 + \hat{\Omega}(n)]
\end{equation}
$\omega$ is represented as  $\omega = \omega_R - i \omega_I$ and the lowest quasi normal modes $\omega(0)$ of the Born-Infeld black holes are computed. \\

First the quasi normal modes are computed to see the behavior with the non-linear parameter  $\beta$ as follows;
\begin{center}
\begin{tabular}{|l|l|l|l|l|r} \hline \hline
 $\beta$ & $\omega_R$  &  $\omega_I$ \\ \hline
0.01 & 0.252195 & 0.035046  \\ \hline
0.03 & 0.258583 & 0.037396  \\ \hline
0.04 & 0.258527 & 0.035400  \\ \hline
0.07 & 0.257616 & 0.034988  \\ \hline
0.1 & 0.256993 & 0.035387  \\ \hline
0.3 & 0.256105 & 0.035558 \\ \hline
0.5 & 0.256011 & 0.035542  \\ \hline

\end{tabular}

\vspace{0.3cm}

\scalebox{.9}{\includegraphics{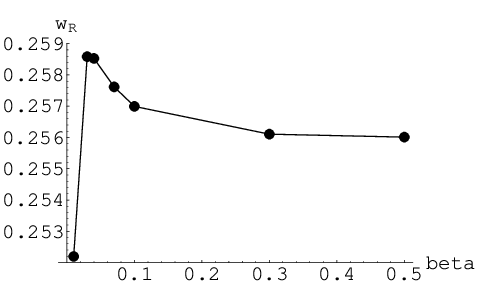}}

\vspace{0.3cm}

\end{center}
Figure 3. The behavior of Re $\omega$ with the non-linear parameter $\beta$ for $M=2$, $Q=1$ and $l=2$.

\begin{center}
\scalebox{.9}{\includegraphics{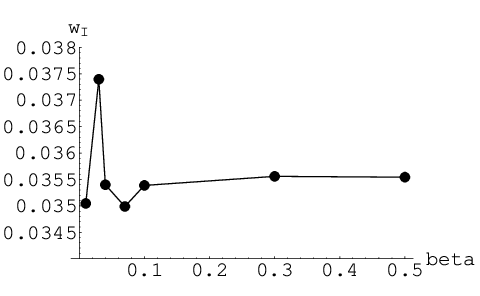}}

\vspace{0.3cm}

\end{center}
Figure 4. The behavior of Im $\omega$ with the non-linear parameter $\beta$ for $M=2$, $Q=1$ and $l=2$

\vspace{0.2cm}

The behavior of the quasi normal modes with varying charge $Q$ was also studied as given in the following table.
\begin{center}
\begin{tabular}{|l|l|l|r} \hline \hline
 $Q$ & $\omega_R$  &  $\omega_I$ \\ \hline
0.2 & 0.481698 & 0.068375 \\ \hline
0.4 & 0.498791 & 0.069957 \\ \hline
0.6 & 0.530753 & 0.072592 \\ \hline
0.7 & 0.554740 & 0.074309 \\ \hline
0.8 & 0.586874 & 0.076306 \\ \hline
0.9 & 0.631678 & 0.078624 \\ \hline
1 & 0.699423 & 0.081372 \\ \hline

\end{tabular}

\vspace{0.3cm}

\scalebox{.9}{\includegraphics{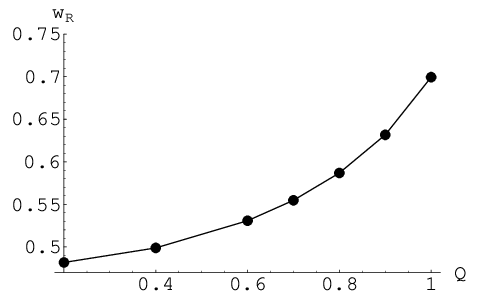}}

\vspace{0.3cm}

\end{center}
Figure 5. The behavior of Re $\omega$ with the charge $Q$ for $M=1$, $\beta=0.4$ and $l=2$

\begin{center}
\scalebox{.9}{\includegraphics{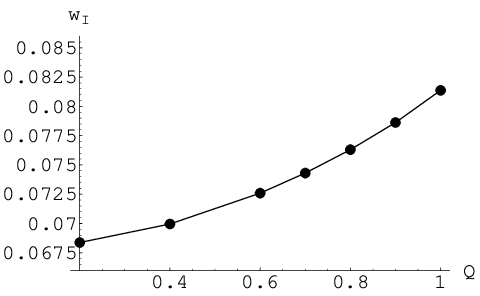}}

\vspace{0.3cm}

\end{center}
Figure 6. The behavior of Im $\omega$ with the charge $Q$ for $M=1$, $\beta=0.4$ and
$l=2$
\\

The behavior of the quasi normal modes of the Born-Infeld  black hole is compared with the Reissner-Nordstrom black hole. The effective potential for the axial perturbation for the Reissner-Nordstrom black hole is computed in Chandrasekhar's book \cite{chandra} which is given below.

\begin{equation}
V_{i} =  \frac{ ( r^2 - 2 Mr + Q^2 )}{r^5} \left[ ( \mu^2 + 2 ) r - q_j \left( 1 + \frac{q_i}{\mu^2 r} \right) \right], \hspace{0.5cm} (i,j=1,2, i \neq j)
\end{equation}
Here,
\begin{equation}
q_1 =  3 M + \sqrt{ 9 M^2 + 4 Q^2 \mu^2}; \hspace{0.5 cm} q_2 =  3 M - \sqrt{ 9 M^2 + 4 Q^2 \mu^2}
\end{equation}
The QNM's for $V_1$ is computed for varying charge and given below.

\begin{center}

\scalebox{.9}{\includegraphics{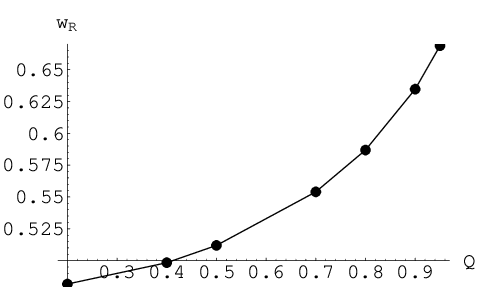}}

\vspace{0.3cm}

\end{center}

Figure 7. The behavior of Re $\omega$ with the charge $Q$ for $M=1$ and $l=2$ for the Reissner-Nordsrom black hole

\begin{center}
\scalebox{.9}{\includegraphics{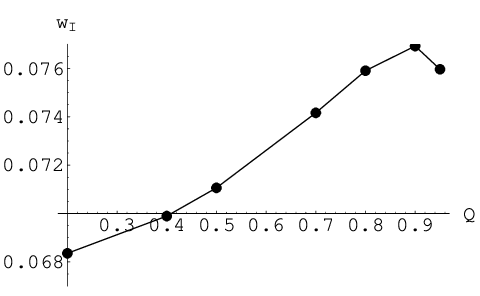}}

\vspace{0.3cm}

\end{center}

Figure 8. The behavior of Im $\omega$ with the charge $Q$ for $M=1$ and $l=2$ for the Reissner-Nordsrom black hole
\\

It is interesting to note that when the charge increases, the imaginary part of the QNM's continue to increase for the Born-Infeld black hole while for the Reissner-Nordstrom black hole  it reaches a maximum and decreases.
\\

The behavior of the quasi normal modes with spherical index $l$ is given in the following table.
\begin{center}
\begin{tabular}{|l|l|l|r} \hline \hline
 $l$ & $\omega_R$  &  $\omega_I$ \\ \hline
2 & 0.255970 & 0.035534 \\ \hline
3 & 0.359628 & 0.039847 \\ \hline
4 & 0.462125 & 0.042062 \\ \hline
5 & 0.564053 & 0.043404 \\ \hline
6 & 0.665646 & 0.044302 \\ \hline
7 & 0.767021 & 0.044943 \\ \hline
\end{tabular}

\vspace{0.3cm}

\scalebox{.9}{\includegraphics{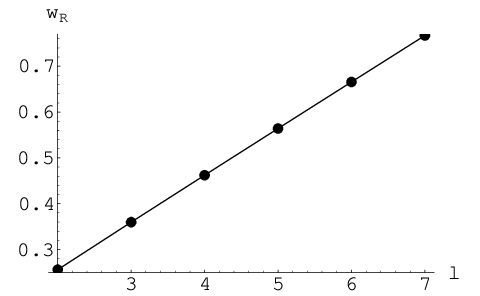}}

\vspace{0.3cm}
\end{center}
Figure 9. The behavior of Re $\omega$ with the the spherical index $l$ for $M=2$, $Q=1$ and $\beta=1$

\begin{center}
\scalebox{.9}{\includegraphics{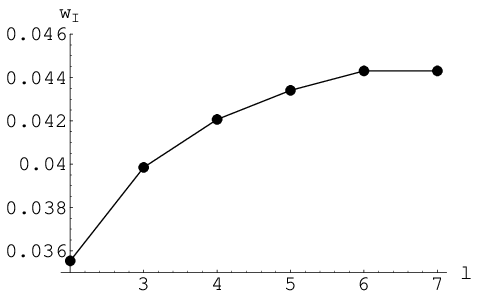}}

\vspace{0.3cm}

\end{center}
Figure 10. The behavior of Im $\omega$ with the spherical index $l$ for $M=2$, $Q=1$ and $\beta=1$


\section{Conclusion}

The gravitational perturbations of charged  black holes in Einstein-Born-Infeld gravity are studied. One dimensional Schrodinger type wave equations are  derived for the axial perturbations. The behavior of the effective potential with the non-linear parameter $\beta$ and charge $Q$ are observed. From the behavior of the potentials it is concluded that the black holes are stable classically.

The lowest quasi-normal modes  are computed  using the WKB method.  It was observed that when  $\beta$ increases, the real part of QNM's increases to a maximum and decreases to a stable value. On the other hand, the imaginary part of QNM's shows an oscillating behavior before it stabilizers to a fixed value. Such behavior indicates  that QNM's depend on $\beta$ in a non trivial manner. As explained in \cite{rasheed} and \cite{nora},  the number of horizons of the black hole depend on $\beta$ for the same charge and  mass. It is interesting to study how QNM's behavior changes depending on the horizon structure.

It was also observed that when charge increases,  QNM's increases. In comparison, for the Reissner-Nordstrom black hole the QNM's increases with the charge and decreases when the extremality reaches. I have not done a close study as to how the extreme nature of Born-Infeld black holes effects the behavior of the QNM's. This would be   left for future work.

It is also noted that when the spherical index $l$ is increased,  Re $\omega$ increases leading to greater oscillations. On the other hand Im $\omega$ approaches to a fixed value for larger $l$.

There are several avenues to proceed from here. One of the important extensions of this work is to compute the potentials and QNM's for the polar perturbations of the Born-Infeld black hole. It is well known that for the Schwarzschild and Reissner-Nordstrom black hole, the two potentials for axial and perturbations are related to each other in a simple manner leading to equality in reflexion and transmission coefficients. Hence they also have the same QNM's for both types of perturbations. Our expectation is that the modes would be  {\it isospectral } for the Born-Infeld black hole as well. However, for the charged dilaton black hole, it was shown that the presence of the dilaton in fact breaks isospectrality \cite{fera}.

It would be   interesting to investigate the supersymmetric nature of the Born-Infeld black hole discussed in this paper. It is a well known fact that the extreme Reissner-Nordstrom  black hole can be embedded in $N=2$ supergravity theory \cite{gib2} \cite{gib3}. Onozawa  et.al. \cite{ono} showed that the QNM's of the extreme RN black hole for spin 1, 3/2 and 2 are the same. 

There are other approaches to  calculate QNM's other than the method followed in this paper. For example power series expansion of the wave function is one of the   methods in computing QNM's as  used by Horowitz and Hubeny \cite{horo}. QNM's of   higher modes are obtained with great precision with a semi-analytical method developed by Weaver \cite{wea}. This   has been applied to Schwarzchild and Kerr black hole. It would be interesting to apply other methods to find QNM's and compare the results obtained in this paper. 

From the Loop Quantum Gravity point of view, it is necessary to compute the QNM's for higher overtones. The method used in this paper is not accurate for large $n$. However, there are other approaches applied to compute highly damped QNM's \cite{vit1} \cite{vit2} \cite{vit3} \cite{vit4}.

\vspace{0.5cm}

{\bf Acknowledgments}: I like to thank Don Krug  for helpful discussions. I also like to thank Amy Matracia for reading the manuscript carefully and making suggestions. This work was support in part by a CINSAM grant and a EPSCOR grant.
 


\begin{thebibliography}{99}


\bibitem{born} M. Born and L. Infeld, Proc. Roy. Soc. Lond. {\bf A144} (1934) 425.
\bibitem{leigh} R. G. Leigh, Mod. Phys. Lett {A4} (1989) 2767.
\bibitem{frad} E. S. Fradkin and A. A. Tseytlin, Phys. Lett. {\bf B163} (1985) 123.
\bibitem{tsey} A. A. Tseytlin, Nucl. Phys. {\bf B276} (1986) 391.
\bibitem{gib1} G. W. Gibbons, Rev.Mex.Fis. 49S1:19 (2003)

\bibitem{rasheed} D. A. Rasheed, hep-th/9702087

\bibitem{nora} N. Breton, gr-qc/0109022


\bibitem{kok1} K.D. Kokkotas, B.G. Schmidt, Living Rev. Relativ. {\bf2} (1999) 2
\bibitem{hod} S. Hod, Phys. Rev. Lett. {\bf 81} (1998) 4293

\bibitem{corichi} A. Corichi, Phys. Rev. {\bf D67} (2003) 087502

\bibitem{mot1} L. Motl, Adv. Theor. Math. Phys. {\bf 6} (2003) 1135

\bibitem{dreyer} O. Dreyer, Phys. Rev. Lett. {\bf 90} (2003) 08130

\bibitem{mot2} L. Motl and A. Neitzke, Adv. Theor. Math. Phys. {\bf 7} (2003) 307


\bibitem{van} A. Maassen van den Brink, J. Math. Phys. {\bf 45} (2004) 327

\bibitem{set1} M. Setare, Class. Quant. Grav. {\bf 21} (2004) 1453

\bibitem{set2} M. Setare, Phys. Rev. {\bf D69} (2004) 044016

\bibitem{kun} G. Kunstatter, Phys. Rev. Lett. {\bf 90} (2003) 161301

\bibitem{ander}N. Andersson and C.J. Howls, Class. Quan. Grav. {\bf 21} (2004) 1623


\bibitem{aha} O. Aharony, S.S. Gubser, J. Maldacena, H. Ooguri \& Y. Oz, Phys. Rept. {\bf 323} 183 (2000)


\bibitem{chan} J. Chan \& R. Mann, Phys. Rev. {\bf D55} (1997) 7546; Phys. Rev. {\bf D 59} (1999) 064025
\bibitem{horo} G.T. Horowitz \& V. E. Hubeny, Phys. Rev. {\bf D62} (2000) 024027

\bibitem{car1} V. Cardoso \& J.P.S. Lemos, Phys. Rev. {\bf D64} (2001) 084017

\bibitem{car2} V. Cardoso \& J.P.S. Lemos, Phys. Rev. {\bf D67} (2003) 084020

\bibitem{moss} I.G. Moss \& J.P. Norman, Class. Quan. Grav. {\bf 19} (2002), 2323

\bibitem{wang} B. Wang, C.Y.Lin \& E. Abdalla, Phys. Lett {\bf B481} (2000) 79

\bibitem{kok2} E. Berti \& K.D. Kokkotas, Phys. Rev {\bf D67} (2003) 064020

\bibitem{kon1} R. A. Konoplya, Phys. Rev. {\bf D68} (2003) 024018


\bibitem{kon2} R. A. Konoplya, Phys. Lett. {\bf B550} (2002) 117

\bibitem{kon3} R. A. Konoplya, Phys. Rev. {\bf D66} (2002) 084007

\bibitem{kon4} R. A. Konoplya, Phys. Rev. {\bf D66} (2002) 044009

\bibitem{kon5} R. A. Konoplya, Gen. Rel. Grav. 34 (2002) 329

\bibitem{li} X. Li, J. Hao \& D. Liu, Phys. Lett. {\bf B507} (2001) 312

\bibitem{aros} R. Aros, C. Martinez, R. Troncoso \& J. Zanelli, Phys. Rev. {\bf D67} (2002) 044014

\bibitem{car3} V. Cardoso \& J.P.S. Lemos, Phys. Rev. {\bf D63} (2001) 124015

\bibitem{abd} E. Abdalla, B. Wang, A. Lima-Santos \& W.G. Qiu, Phys. Lett. {\bf B38} (2002) 435


\bibitem{fer1} S. Fernando \& D. Krug, Gen. Rel. Grav. {\bf 35} (2003) 129

\bibitem{dey} T. K. Dey, Phys. Lett. {\bf B 595} 484 (2004)

\bibitem{chandra} ``The Mathematical Theory of Black Holes'', S. Chandrasekhar, Oxford University Press (1992) 

\bibitem{regge} T. Regge \& J. Wheeler, Phys. Rev. {\bf 108} (1957) 1063

\bibitem{zer1} F. Zerilli, Phys. Rev. {\bf D 2} (1971) 2141

\bibitem{zer2} F. Zerilli, Phys. Rev. {\bf D 9} (1974) 860

\bibitem{mon} V. Moncrief, Phys. Rev. {bf D 9} (1974) 2702; {\it ibid}. {\bf D10} (1974) 1057; {\it ibid}. {\bf D 12} (1975) 1526

\bibitem{teuk} S. Teukolsky, Phys. Rev. Lett. {\bf 29} (1972) 1114

\bibitem{wil} C.F.E. Holzhey and F. Wilczek, Nucl. Phys. {\bf B 380} (1992) 447

\bibitem{fera} V. Ferrari, M. Pauri \& F. Piazza, Phys. Rev {\bf D 63} (2001) 064009

\bibitem{bir1} D. Birmingham, Phys. Rev {\bf D64} (2001) 064024

\bibitem{fer2} S. Fernando, Gen. Rel. Grav. {\bf 36}  (2004) 71
\bibitem{nun} A. Nunez \& A. Starinets, Phys. Rev. {\bf D67} (2003) 124013

\bibitem{will} S. Iyer \& C.M. Will, Phys. Rev. {\bf D 35} (1987) 3621

\bibitem{iyer} S. Iyer, Phys. Rrev. {\bf D 35} (1986) 3632

\bibitem{koko1} K. D. Kokotas \& B.F. Shutz Phys. Rev. {\bf D 37} (1988) 3378 


\bibitem{fer3} S. Fernando \& K. Arnold, Gen. Rel. Grav. {\bf 36} (2004)  1805-1819


\bibitem{gib2} G. W. Gibbons \& C.M. Hull, Phys. Lett. {\bf B109} (1982) 190

\bibitem{gib3} G. Gibbons, {\it in Unified Theories of Elementary Particles, Crtical Assessment and prospects}, proceedings of the Heisenberg Symposium, Munchen, Germany, 1981, edited by P. Breitenlohner and H. P. Durr, Lecture Notes in Physics Vol.160(Springer-Verlag, Berlin 1982)

\bibitem{ono} H. Onozawa, T. Okumura, T. Mishima \& H. Ishihara, Phys. Rev. {\bf D55} (1997) 4529


\bibitem{wea} E. W. Leaver, Proc. R. Soc. {\bf A 402} (1985) 285 


\bibitem{vit1} V. Cardoso, R. Konoplya \& J.P.S. Lemos, Phys. Rev. {\bf D 68} (2003) 044024

\bibitem{vit2} V. Cardoso, J. Natario \& R, Schiappa, Jour.Math.Phys. {\bf 45} 4698 (2004)

\bibitem{vit3} E. Berti, V. Cardoso, K. D. Kokkotas \& H. Onozawa, Phys. Rev. {\bf D 68} (2003) 124018

\bibitem{vit4} E. Berti, V. Cardoso \&  S. Yoshida, Phys. Rev. {\bf D 69} (2004) 124018



\ \end{thebibliography}
\end{document}